\def\la{\hbox{{\lower -2.5pt\hbox{$<$}}\hskip -8pt\raise
-2.5pt\hbox{$\sim$}}}
\def\ga{\hbox{{\lower -2.5pt\hbox{$>$}}\hskip -8pt\raise
-2.5pt\hbox{$\sim$}}}
\def\ltsima{$\; \buildrel < \over \sim \;$}
\def\simlt{\lower.5ex\hbox{\ltsima}}
\def\gtsima{$\; \buildrel > \over \sim \;$}
\def\simgt{\lower.5ex\hbox{\gtsima}}

\documentstyle[epsfig]{elsart}

\newcommand{\dsc}[4]{$#1\pm#2\,(#3)\,$\boldmath{$[#4]$}}

\begin{document}
\begin{frontmatter}
\title{On the statistical significance of the GZK feature in the
spectrum of ultra high energy cosmic rays}

\author[infn]{Daniel De Marco\thanksref{corr1}}
\author[inaf]{Pasquale Blasi\thanksref{corr2}}
\author[uofc,cfcp]{Angela V. Olinto\thanksref{corr3}}
\address[infn]{INFN \& Universit\`a degli Studi di Genova\\ Via
Dodecaneso, 33 -
16100 Genova, ITALY}
\address[inaf]{INAF/Osservatorio Astrofisico di Arcetri\\ Largo E. Fermi, 5 -
50125 Firenze, ITALY}
\address[cfcp]{Center for Cosmological Physics \\ The University of Chicago,
Chicago, IL 60637, USA}
\address[uofc]{Department of Astronomy \& Astrophysics,  \& Enrico Fermi
Institute, \\ The University of Chicago, Chicago, IL 60637, USA}
\thanks[corr1]{E-mail: ddm@ge.infn.it}
\thanks[corr2]{E-mail: blasi@arcetri.astro.it}
\thanks[corr3]{E-mail: olinto@oddjob.uchicago.edu}

\begin{abstract}
The nature of the unknown sources of ultra-high energy cosmic rays can be
revealed through the detection of the GZK feature in the cosmic ray
spectrum, resulting from the production of pions by ultra-high
energy protons scattering off the cosmic microwave background. Here we show
that the GZK feature cannot be accurately determined  with the small
sample of events with energies $\sim 10^{20}$ eV  detected thus far
by the largest two experiments, AGASA and HiRes.   With the
help of numerical simulations for the propagation of cosmic rays,
we find the error bars around the GZK feature are dominated by
fluctuations which leave a determination of the GZK feature unattainable
at present. In addition, differing results from AGASA and HiRes suggest the
presence of $\sim 30\%$ systematic errors that may be due to discrepancies in
the relative energy determination of the two experiments. Correcting for these
systematics, the two experiments are brought into agreement at energies below
$\sim 10^{20}$ eV.   After simulating the GZK feature for
many realizations and different injection spectra, we determine the best
fit injection spectrum required to explain  the observed spectra at
energies above $10^{18.5}$ eV. We show that the discrepancy between the
two experiments at the highest energies has low statistical significance
(at the  2 $\sigma$ level) and that the corrected spectra are best fit by
an injection spectrum with spectral index $\sim 2.6$.  Our results clearly show
the need for much larger experiments such as Auger, EUSO, and OWL, that
can increase the number of detected events by 2 orders of magnitude. Only
large statistics experiments can finally prove or disprove the  existence
of the GZK feature in the cosmic ray spectrum.
\end{abstract}
\end{frontmatter}

\section{Introduction}

The presence or lack of a feature in the spectrum of ultra-high energy
cosmic rays (UHECRs) is key in determining the nature of
these particles and their sources.
Astrophysical proton  sources  distributed homogeneously in the universe
produce a feature in the spectrum due to the production of pions off the
cosmic microwave background. This feature, consisting of a rather sharp
suppression of the flux, occurs at energies above $7 \times 10^{19}$ eV,
as a result of the threshold in
the production of pions in the final state of a proton-photon inelastic
interaction. This important prediction was made  independently by Greisen and
by Zatsepin and Kuzmin \cite{GZK66}. The resulting spectral feature is now
known as the GZK cutoff (or  feature, as we prefer to call it).
Alternative models for UHECR sources that involve new physical processes may
evade the presence of this feature (see, e.g., \cite{bbv98}).
Recent reviews on the origin and propagation of the
ultra-high energy  cosmic rays can be found in \cite{BS00,Olin00}, while a
recent review  of the observations can be found in \cite{watson}.

The detection of cosmic ray events with energy above $E_{GZK}\sim 7\times
10^{19}$  eV does not necessarily imply that the GZK feature is not present:
what  characterizes the presence of the GZK feature is the relative number of
events  above and below $E_{GZK}$ when both sides of the spectrum can be
accurately determined.  The steep injection spectra required to fit the
observations below $E_{GZK}$ imply that only a handfull of events above
$10^{20}$ eV can be detected during the operation time of experiments such as
AGASA and HiRes. This makes the identification of the GZK  feature by these
experiments extremely difficult. The problem is  exacerbated by the
fluctuations due to the discreteness of the process of photo-pion production,
as will be discussed below. These uncertainties need to be considered when
attempting a determination of the best fit injection spectrum  of the
particles, and in order to quantify the statistical significance of the
presence or absence of the GZK feature in the observed spectrum.

The discrepancy between the results of the two largest
experiments has generated much debate. AGASA \cite{AGASA} reports a
higher number of events above $E_{GZK}$ than expected while
HiRes\cite{hiresprop,HIRES1,HIRES2} reports a flux consistent with
the GZK feature.  Here we investigate in detail the statistical
significance of this discrepancy as well as the significance of the
presence or absence of the GZK feature in the data. We find that neither
experiment has the necesseray statistics to establish if the spectrum of
UHECRs has a GZK feature.  In addition, a systematic error in the energy
determination of the two experiments seems to be required in order to
make the two sets of observations compatible in the low energy range,
$10^{18.5}-10^{19.6}$ eV, where enough events have been
detected to make the measurements reliable. The combined
systematic errors in the energy determination is likely to be
$\sim 30\%$. If we decrease the  AGASA energies by  $15\%$
while increasing  HiRes energies also by $15\%$, the two experiments predict
compatible fluxes at energies below $E_{GZK}$ and at energies above
$E_{GZK}$ the fluxes are within $\sim 2\sigma$ of each other. In this case,
the best fit injection spectrum has a spectral index of $\sim 2.6$ but a
determination of the GZK feature has very low significance.
The detection or non-detection of the GZK feature in the
cosmic ray spectrum remains open to investigation by
future generation experiments, such as the Pierre Auger project
\cite{auger} and the EUSO \cite{EUSO} and OWL\cite{OWL} experiments.

This paper is planned as follows: in \S 2 we describe our simulations.
In \S 3 we illustrate the present observational situation, limiting
ourselves to AGASA and HiResI, and compare the data to the predictions
of our simulations. We conclude in \S 4.

\section{UHECR spectrum simulations}

We assume that ultra-high energy cosmic rays are protons injected with a
power-law spectrum  in extragalactic sources. The injection spectrum is
taken to be of the form
\begin{equation}
F(E) d E = \alpha E^{-\gamma} \exp(-E/E_{\rm max}) d E
\label{eq:inj}
\end{equation}
where $\gamma$ is the spectral index, $\alpha$ is
a normalization constant, and $E_{\rm max}$ is the maximum energy at the
source. Here we fix $E_{\rm max} =10^{21.5}$ eV, large enough
not to affect the statistics at much lower energies. As shown in
\cite{Blanton:2000dr}, the induced spectrum of a uniform distribution of
sources in space is almost indistinguishable from a distribution with the
observed  large scale structure in the galaxy distribution. Based on this
result, we assume a spatially uniform  distribution of sources and do not
take into account luminosity evolution in order to avoid the introduction
of additional parameters.

We simulate the propagation of protons from source to observer by including the
photo-pion production, pair production, and adiabatic energy losses due to
the expansion of the universe.
In each step of the simulation, we calculate the pair production losses
using the continuous energy loss approximation given the small inelasticity in
pair production ($2 m_{\rm e}/m_{\rm p}\simeq10^{-3}$).  For the rate of
energy loss due to pair production at redshift $z=0$, $\beta_{\rm pp}(E,z=0)$,
we use the results from \cite{Blumenthal:nn,czs}. At a given reshift
$z>0$,
\begin{equation}
\beta_{\rm pp}(E,z)=(1+z)^3 \beta_{\rm pp}((1+z)E, z=0)\,.
\end{equation}
Similarly, the rate of adiabatic energy losses due to redshift is
calculated in each step  using
\begin{equation}
\beta_{\rm rsh}(E,z)=H_0 \left[\Omega_M (1+z)^3 +
\Omega_\Lambda\right]^{1/2}
\end{equation}
with $H_0=75 ~ {\rm km~ s^{-1} Mpc}^{-1}$.

The photo-pion production is simulated in a way similar to that described
in ref.~\cite{Blanton:2000dr}. In each step, we first calculate the
average number of photons able to interact via photo-pion production
through the expression:
\begin{equation}
\langle N_{\rm ph}(E,\Delta s) \rangle=\frac{\Delta s}{l(E, z)},
\end{equation}
where $l(E,z)$ is the interaction length for photo-pion production of a
proton with energy $E$ at redshift $z$ and $\Delta s$ is a step size,
chosen to be much smaller than the interaction length (typically we
choose $\Delta s=100 ~{\rm kpc}/(1+z)^3$).

\begin{figure}
\begin{center}
\includegraphics[width=0.7\textwidth]{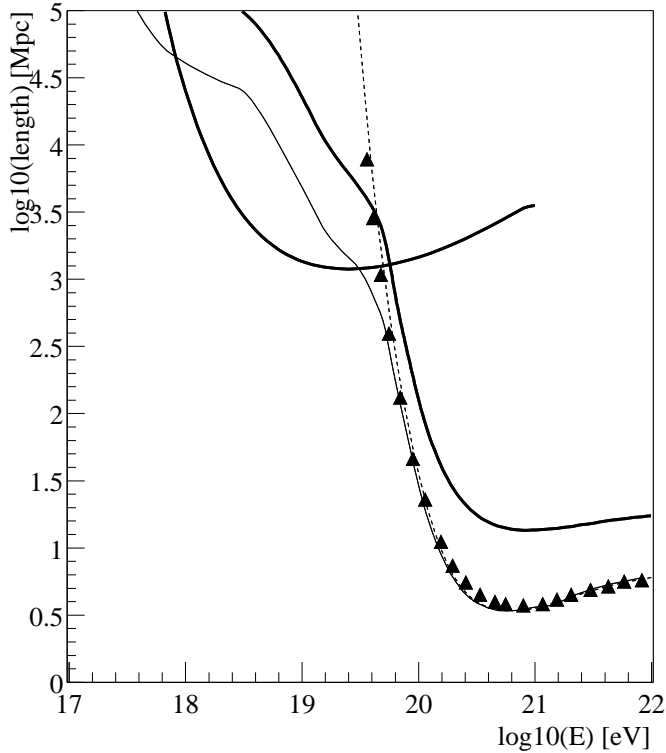}
\caption{Interaction length for photopion production as calculated in this
paper (dashed line) compared to the interaction length of \cite{BS00}
(solid thin line) and of \cite{Stanev} (triangles). The thick solid lines
are the loss lengths for photopion production (on the right) and of pair
production (on the left).}
\label{fig:il}
\end{center}
\end{figure}

In Fig. \ref{fig:il} we plot the interaction length for photopion production
used in \cite{BS00} (solid thin line), and in \cite{Stanev} (triangles).
The dashed line is the result of our calculations (see below),
which is in perfect agreement with the results of \cite{BS00,Stanev}.
The apparent discrepancy at energies below $10^{19.5}$ eV with the prediction
of Ref. \cite{BS00} is only due to the fact that we consider only microwave
photons as background, while in \cite{BS00} the infrared background was also
considered. For our purposes, this difference is irrelevant as can be seen
from  the loss lengths plotted in Fig. \ref{fig:il}.
The rightmost thick solid line is the loss length for photopion production
\cite{BS00}, while the other thick solid line is the loss length for
pair production. In the present calculations, we do not use the loss
length of photopion production which is related to the interaction length
through an angle  averaged inelasticity. Unlike what was done in
\cite{Blanton:2000dr}, in the current simulations we evaluate the
inelasticity for each  single proton-photon scattering using the kinematics,
rather than adopting  an angle averaged value.

We calculate the interaction length, $l(E)$, as:
\begin{equation}
l(E)^{-1}=\int{\rm d}\varepsilon n(\varepsilon)\int^{+1}_{-1}
{\rm d}\mu\frac{1-\mu\beta}{2}\sigma(s)
\end{equation}
where
$n(\varepsilon)$ is the number density of the CMB photons
per unit energy at energy $\varepsilon$, $\beta$ is the velocity of the proton,
$\mu$ is the cosine of the interaction angle, and $\sigma(s)$ is the total
cross section for photo-pion production for the squared center of mass energy
\begin{equation}
s=m^2+2\varepsilon E\left(1-\mu\beta\right)\,.
\end{equation}

For the calculation of the interaction length we adopt the ${\rm p}+\gamma
\rightarrow{\rm hadrons}$ cross section given in \cite{pdgcrosssec},
considering only the photons of the cosmic microwave background as
targets. The calculated interaction length (see dashed line in
Fig.~\ref{fig:il}) is in good agreement with the interaction length
calculated in
\cite{BS00} and in \cite{Stanev}.

Once the interaction length is known, we then sample a Poisson
distribution with mean $\langle N_{\rm ph}(E,\Delta s) \rangle$, to determine
the  actual number of photons encountered during the step $\Delta s$. When a
photo-pion interaction occurs, the energy $\epsilon$ of the photon is extracted
from the Planck distribution, $n_{ph}(\epsilon,T(z))$, with temperature
$T(z)=T_0 (1+z)$, where $T_0=2.728$ K is the temperature
of the cosmic microwave background at present. Since the microwave photons
are isotropically distributed, the interaction angle, $\theta$, between the
proton and the photon is sampled randomly from a distribution which is flat
in $\mu={\rm cos}\theta$. Clearly only the values of $\epsilon$ and
$\theta$ that generate a center of mass energy above the threshold for
pion production are considered. The energy of the proton in the final
state is calculated at each interaction from kinematics.
The simulation is carried out until the statistics of events
detected above some energy reproduces the experimental numbers.
By normalizing the simulated flux by the number of events above an energy
where experiments have high statistics, we can then ask what are the
fluctuations in numbers of events above a higher energy where
experimental results are sparse. The fits are therefore most sensitive
to the energy regions below $E_{GZK}$ and give a good estimate of the
uncertainties in the present experiments for energies above  $E_{GZK}$.
In this way we have a direct handle on the fluctuations that can be
expected in the observed flux due to the stochastic nature of photo-pion
production and to cosmic variance.

The simulation proceeds in the following way: a source distance is generated
at random from a uniform distribution in a universe with $\Omega_\Lambda=0.7$
and $\Omega_m=0.3$. In a Euclidean universe, the flux from a source would
scale as $r^{-2}$ where $r$ is the distance between the source and the
observer. On the other hand, the number of sources between $r$ and $r+dr$ would
scale as $r^2$, so that the probability that a given event has been
generated by a source at distance $r$ is independent of $r$: sources
at different distances have the same probability of generating any
given event. In a flat universe with a cosmological constant, this is
still true provided the distance $r$ is taken to be
\begin{equation}
r = c \int_{t_g}^{t_0} \frac{dt}{R(t)},
\end{equation}
where $t_g$ is the age of the universe when the event was generated, $t_0$
is the present age of the universe, and $R(t)$ is the scale factor of the
universe.
\begin{figure}
\begin{center}
\includegraphics[width=0.9\textwidth]{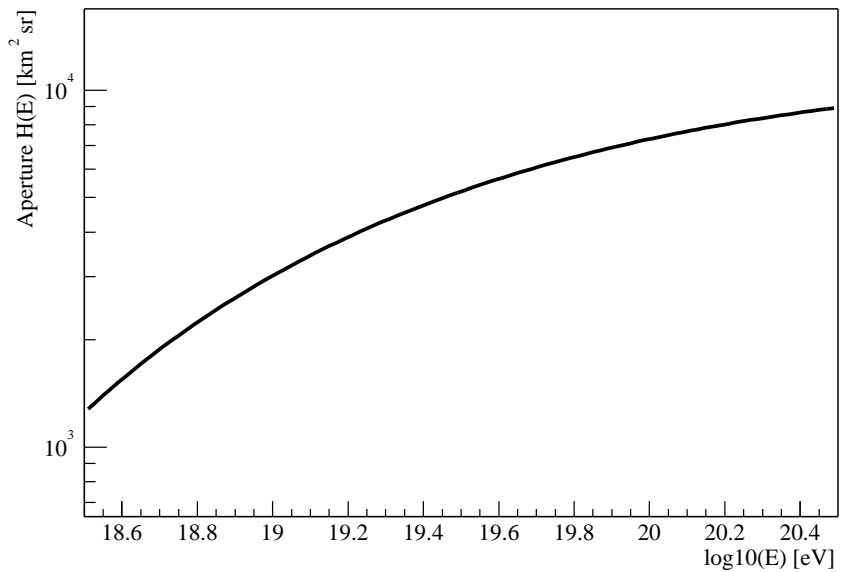}
\caption{Aperture of the HiResI experiment as a function of the energy from
\cite{HIRES2}.}
\label{fig:hiresexp}
\end{center}
\end{figure}
Once a source distance has been selected at random,
a particle energy is also assigned from a distribution that reflects
the injection spectrum, chosen as in Eq. (\ref{eq:inj}). This particle
is then propagated to the observer and its energy recorded. This procedure
is repeated until the number of events above a threshold energy, $E_{th}$
is reproduced. With this procedure we can assess the significance of
results from present experiments with limited statistics of events.
There is an additional complication in that
the aperture of the experiment usually depends on energy. This is taken into
account by allowing the event to be detected or not detected depending
upon the function $H(E)$ that describes the energy dependence of the
aperture.

We only study the spectrum above $10^{18.5}$ eV where the flux is
supposed to be dominated by  extragalactic sources. For this energy range,
we focus on the experiments that have the best statistics: AGASA and
HiResI.  For the AGASA experiment the exposure
is basically flat above $10^{19}$ eV,  while for HiRes the exposure is as
plotted in  Fig. (\ref{fig:hiresexp}) \cite{HIRES2}. For AGASA data, the
simulation is stopped when the number of events above $E_{th} = 10^{19}$ eV
equals 866.  For HiRes this number is 300. Note that while for AGASA the
number of detected events actually corresponds to the generated events, for
HiRes the number of detected events requires a correction due to the
energy dependence of the aperture $H(E)$. This correction allows one
to reconstruct the observed spectrum. The statistical error in the
energy determination is accounted for in our simulation by generating
a {\it detection} energy chosen at random from a Gaussian distribution
centered at the arrival energy $E$ and with width $\Delta E/E=30\%$
for both experiments.

\begin{figure}
\begin{center}
\includegraphics[width=0.8\textwidth]{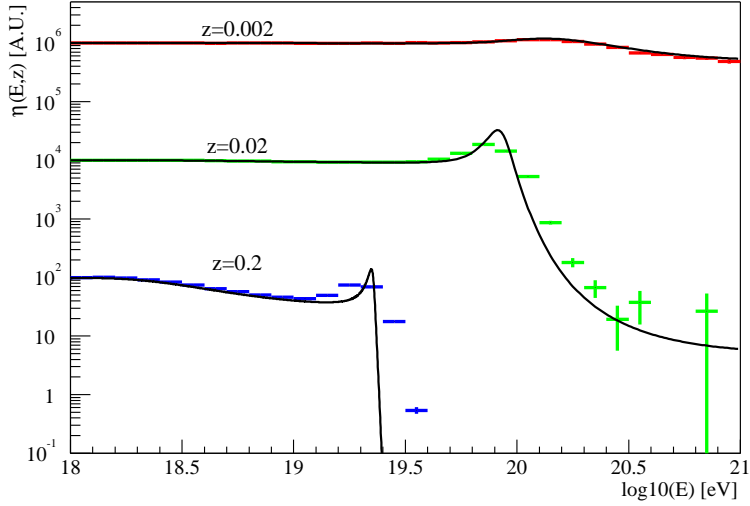}
\caption{Comparison between analytical calculations and the results of our
simulation for the modification factor, for injection spectrum $E^{-2.1}$
\cite{beregrig} and three values of the source redshift ($z=0.002$, $z=0.02$
and $z=0.2$ from top to bottom).}
\label{fig:an}
\end{center}
\end{figure}

Our simulations reproduce well the predictions of analytical calculations,
in particular at the energies where energy losses may be approximated as
continuous. In Fig. \ref{fig:an}, we compare the results of our simulation
with analytical calculations carried out as in ref. \cite{beregrig}. The
curves plotted in the figure are the so called modification factors, defined in
\cite{beregrig,berebook} for three different values of the source
redshift ($z=0.002$, $z=0.02$ and $z=0.2$ from top to bottom).
The differential injection spectrum is taken
to be a power law $E^{-2.1}$. The points with error bars are the results
of our simulations with $2\times 10^6$ particles produced
by sources at the redshifts given above. The agreement between the simulated
and analytical calculations at low energies is clear. At the energies
where photo-pion production becomes important, simulations predict a
slightly larger flux than analytical calculations. This is a well known
result, and is due to the discreteness of photo-pion energy losses, that
allow some particles to reach the detector without appreciable losses.

\begin{figure}
\begin{center}
\includegraphics[width=0.8\textwidth]{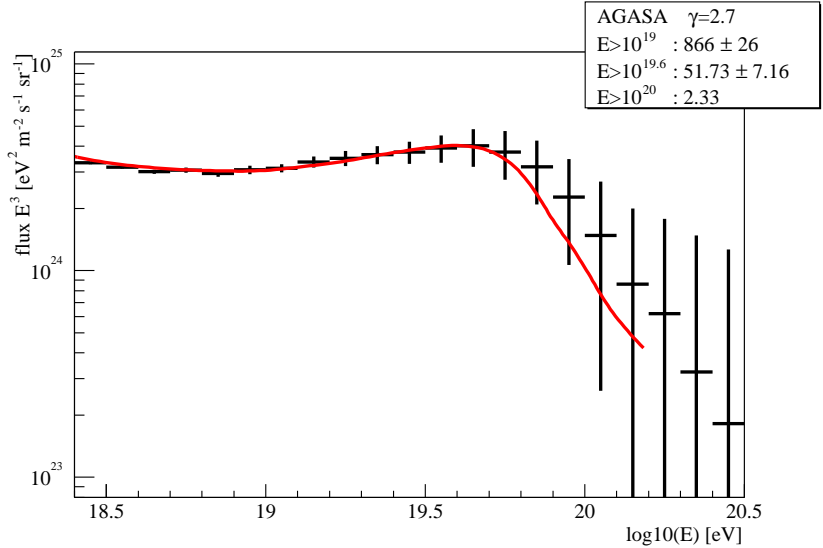}
\caption{Comparison between the results of our simulation and the analytical
calculations of \cite{bereAGN,bgg2002} for the case of astrophysical sources
injecting cosmic rays with a spectrum $E^{-2.7}$ and no luminosity evolution.}
\label{fig:an1}
\end{center}
\end{figure}

In Fig. \ref{fig:an1} we compare the results of our simulations (points with
error bars) with analytical calculations of the diffuse flux of UHECRs
(continuous line) expected if sources with no luminosity evolution inject
UHECRs with a spectrum $E^{-2.7}$ \cite{bereAGN,bgg2002}. The agreement is
evident.

\section{AGASA versus HiResI}

\begin{figure}
\begin{center}
\includegraphics[width=0.9\textwidth]{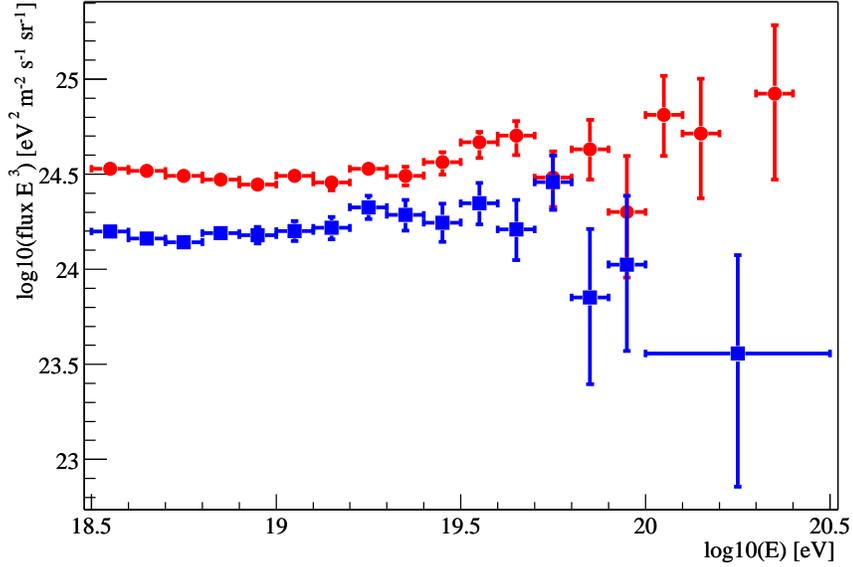}
\caption{Circles show the AGASA spectrum from \cite{AGASA} while squares show
the  HiResI spectrum from
\cite{HIRES2}.}
\label{fig:agasahires}
\end{center}
\end{figure}

The two largest experiments that measured the flux of UHECRs report
apparently conflicting results. The data of AGASA and HiResI on the
flux of UHECRs multiplied as usual by the third power of the energy
are plotted in Fig. \ref{fig:agasahires} (the squares are the HiResI
data while the circles are the AGASA data).
The figure shows that HiResI data are systematically below
AGASA data and that HiResI sees a suppression at $\sim 10^{20}$ eV
that resembles the GZK feature while AGASA does not.

We apply our simulations here to the statistics of events of both
AGASA and HiResI in order to understand whether the discrepancy is
statistically significant and whether the GZK feature has indeed
been detected in the cosmic ray spectrum.
The number of events above $10^{19}$ eV, $10^{19.6}$ eV and
$10^{20}$ eV for AGASA and HiResI are shown in Table
\ref{tab:data_shifted}.
For reasons that will be clear below, we also show in the same
table the number of events above the same energy thresholds,
in the case that AGASA has a systematic error that overestimates the energy
by $15\%$ while HiResI systematically underestimates the energy by $15\%$.
\begin{table}
\begin{center}
\caption{Number of events for AGASA and HiResI detected above the
energy thresholds
reported in the first column.}
\begin{tabular}{c|cc|cc}
$\log(E_{th})\,({\rm eV})$ &
\multicolumn{2}{c}{AGASA} &
\multicolumn{2}{|c}{HiResI} \\
& data & -15\% & data & +15\% \\
\hline
19  & 866 & 651 & 300 & 367\\
19.6 & 72 & 48 & 27 & 39\\
20 & 11 & 7 & 1 & 2.2\\
\hline
\end{tabular}
\label{tab:data_shifted}
\end{center}
\end{table}
In order to understand the difference, if any, between AGASA and HiRes
data we first determine the injection spectrum required to
best fit the observations. In order to do this, we run 400 realizations
of the AGASA and HiRes event statistics, as reported in the column
labelled  {\it data} in Table \ref{tab:data_shifted}.

The injection spectrum is taken to be a power law with index $\gamma$
between 2.3 and 2.9 with steps of 0.1. For each injection spectrum we
calculated the $\chi^2$ indicator (averaged over 400 realizations for each
injection spectrum). The errors used for the evaluation of the $\chi^2$ are
due  to the square roots of the number of observed events.
As reported in Table \ref{tab:chiq}, the best fit injection
spectrum can change appreciably depending on the minimum energy above which
the fit is calculated. In these tables we
give  $\chi_{e}^2(N)$, where $N$ is the number of degrees of  freedom,
while the subscript, $e$, is the logarithm of $E_{th}$ (in eV), the energy
above which the fit has been calculated. The numbers in bold-face correspond
to the best fit injection spectrum.  These fits are dominated by the low
energy data rather than by the poorer statistics at the higher energies.

\begin{table}
\begin{center}
\caption{$\chi^2$ for fits to AGASA and HiResI data above $10^{18.5}$ eV,
$10^{19}$ eV, and $10^{19.6}$ eV for varying spectral index $\gamma$. In
parenthesis the number of degrees of freedom.}\label{tab:chiq}
\begin{tabular}{l|ccc|ccc}
&\multicolumn{3}{c|}{AGASA}&\multicolumn{3}{c}{HiResI}\\
$\gamma$    &   $\chi^2_{18.5}(17)$ &   $\chi^2_{19}(12)$   &
$\chi^2_{19.6}(6)$  &   $\chi^2_{18.5}(15)$ &   $\chi^2_{19}(10)$   &
$\chi^2_{19.6}(4)$\\
\hline
2.3 & 1694 & 34       & 17          & 145     & 29      & 23\\
2.4 & 1215 & 24       & 12          & 80      & 19      & 15\\
2.5 & 724  & 19       & {\bf 10}    & 36      & 14      & 11\\
2.6 & 248  & {\bf 16} & {\bf 10}    & {\bf 14}& 9       & 7\\
2.7 & 82   & 17       & 11          & 33      & {\bf 7} & 5\\
2.8 & {\bf 80}& 21    & 13          & 126     & {\bf 7} & {\bf 4}\\
2.9 & 316  & 27       & 15          & 257     & 9       & {\bf 4}\\
\hline
\end{tabular}
\end{center}
\end{table}

If the data at energies above $10^{18.5}$ eV are taken into account
for both experiments, the best fit spectra are $E^{-2.8}$ for AGASA and
$E^{-2.6}$ for HiRes. If the data at energies above $10^{19}$ eV are
used for the fit, the best fit injection spectrum is $E^{-2.6}$ for
AGASA and between $E^{-2.7}$ and $E^{-2.8}$ for HiRes. If the fit is
carried out on the highest energy data ($E>10^{19.6}$ eV), AGASA
prefers an injection spectrum between $E^{-2.5}$ and $E^{-2.6}$, while
$E^{-2.8}$ or $E^{-2.9}$ fit better the HiRes data in the same energy
region. Note that the two sets of data uncorrected for any possible
systematic errors require different injection spectra that change with
$E_{th}$.

\begin{table}
\begin{center}
\caption{Number of events expected above $E_{th}$(eV) for different
injection spectra assuming the AGASA statistics above $10^{19}$ eV.
In parenthesis are the number of standard deviations, $\sigma$, between the
observed and expected number of events.
In square brackets are the discrepancies calculated
with a combined error bar of simulation and observation uncertainties,
$\sigma_{tot}$.}
\begin{tabular}{c|c|c|c}
${E_{th}}$& $\gamma=2.5$ & $\gamma=2.6$ & $\gamma=2.8$\\
\hline
$10^{19.6}$ & \dsc{65}{8.2}{+0.5}{+0.3} &  \dsc{58}{7.6}{+1.4}{+1.0} &
\dsc{46}{6.8}{+2.8}{+2.2} \\ $10^{20}$   & \dsc{3.5}{1.9}{+2.4}{+2.1} &
\dsc{2.8}{1.7}{+2.6}{+2.3} & \dsc{2.0}{1.4}{+2.8}{+2.6} \\
\hline
\end{tabular}
\label{tab:discrepanza_agasa}
\end{center}
\end{table}

\begin{table}
\begin{center}
\caption{Number of events expected above $E_{th}$(eV) for different
injection spectra assuming the HiResI statistics above $10^{19}$ eV from Table
\ref{tab:data_shifted}. In parenthesis are the number of $\sigma$ between
the observed and  expected number of events.   In square
brackets are the number of $\sigma_{tot}$.}
\begin{tabular}{c|c|c|c}
${E_{th}}$ & $\gamma=2.6$ & $\gamma=2.7$ & $\gamma=2.8$\\
\hline
$10^{19.6}$ & \dsc{31}{5.6}{-0.8}{-0.6} & \dsc{28}{5.3}{-0.2}{-0.1} &
\dsc{26}{5.2}{+0.3}{+0.2} \\ $10^{20}$ & \dsc{1.9}{1.4}{-0.9}{-0.5} &
\dsc{1.5}{1.2}{-0.5}{-0.3} & \dsc{1.3}{1.2}{-0.3}{-0.2} \\
\hline
\end{tabular}
\label{tab:discrepanza_hires}
\end{center}
\end{table}
In order to quantify the significance of the detection or lack of
the GZK flux suppression,  we report in Tables \ref{tab:discrepanza_agasa}
and \ref{tab:discrepanza_hires} the mean number of events above the
indicated energy threshold, $\langle N (E > E_{th})
\rangle$,  for different injection specta. In parenthesis , we show the
  discrepancy between the  observations as in Table
\ref{tab:data_shifted} and the simulations in number of standard
deviations, $\sigma$, where $\sigma^2 = \langle {N(E > E_{th})^2 -
\langle N (E > E_{th}) \rangle^2} \rangle$. From Tables
\ref{tab:discrepanza_agasa} and \ref{tab:discrepanza_hires} we can see
that while HiResI is consistent with the existence of the GZK feature in
the spectrum of UHECRs, AGASA detects an increase in flux, but only
at the $\sim 2.5\sigma$ level for the best fit injection spectra.

\begin{figure}
\begin{center}
\includegraphics[width=0.7\textwidth]{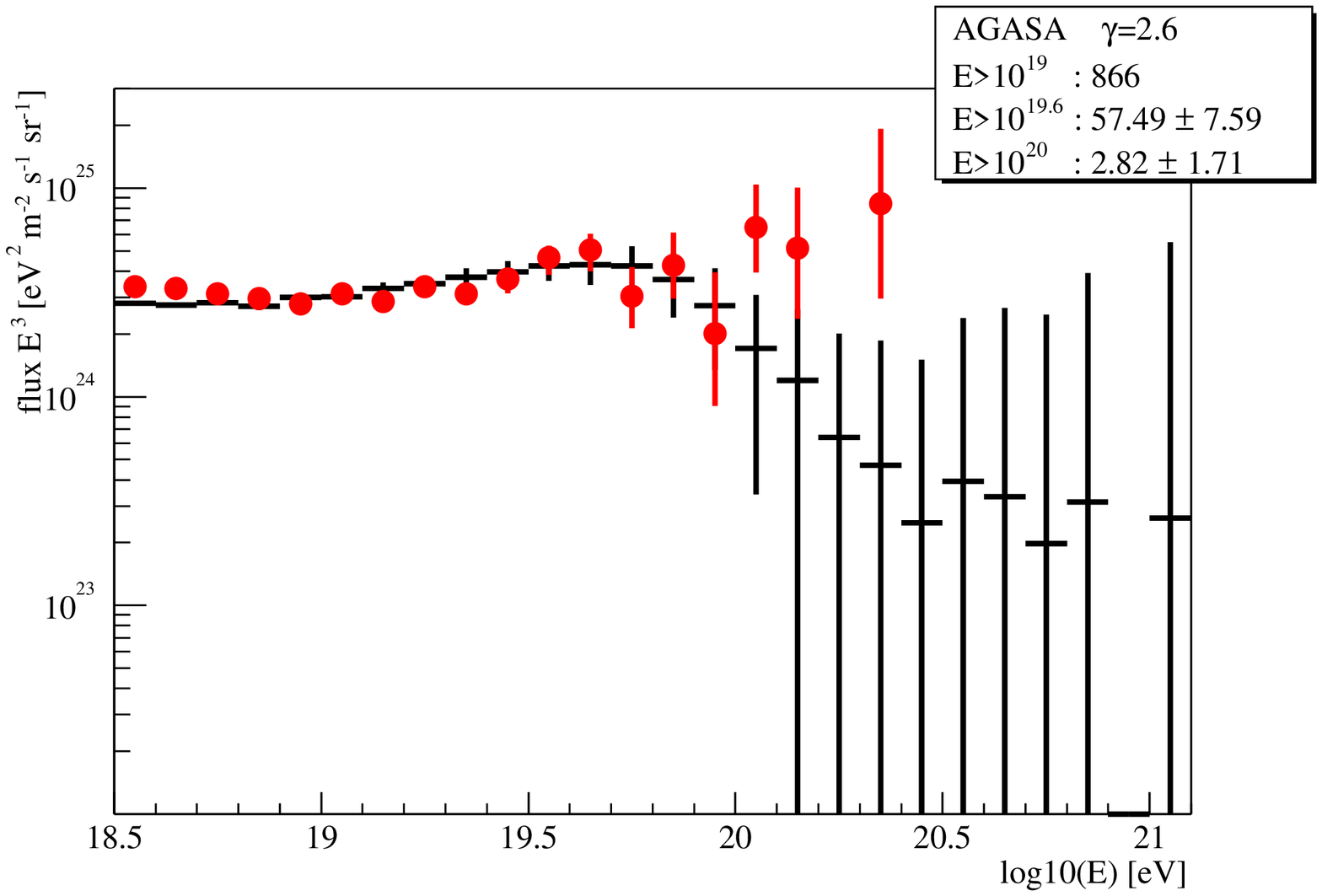}
\includegraphics[width=0.7\textwidth]{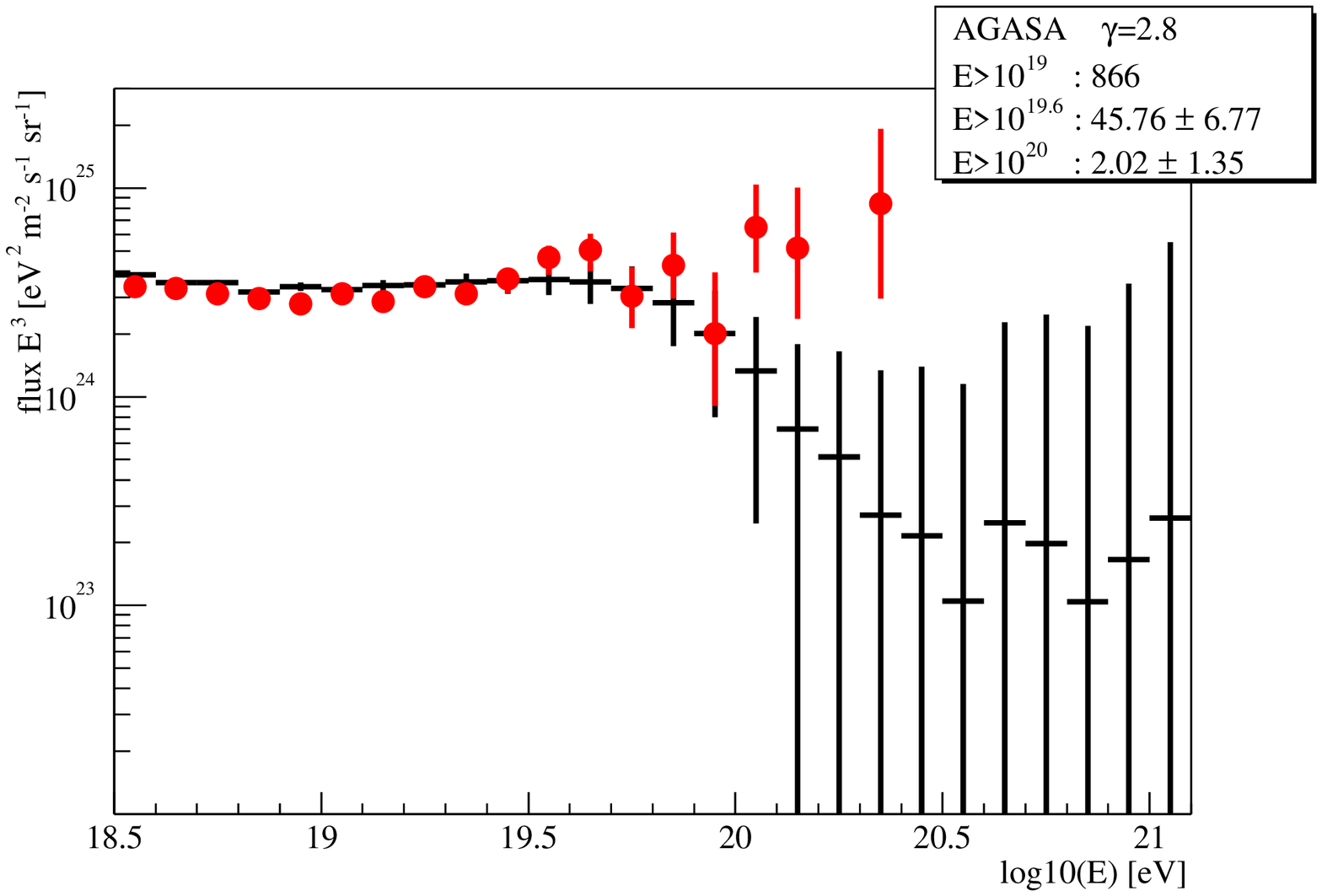}
\caption{Simulations of AGASA statistics with injection spectra
$E^{-2.6}$ (upper plot) and $E^{-2.8}$ (lower plot). The crosses with
error bars are the results of simulations, while the grey
points are the AGASA data.}\label{fig:agasa_2627}
\end{center}
\end{figure}

\begin{figure}
\begin{center}
\includegraphics[width=0.7\textwidth]{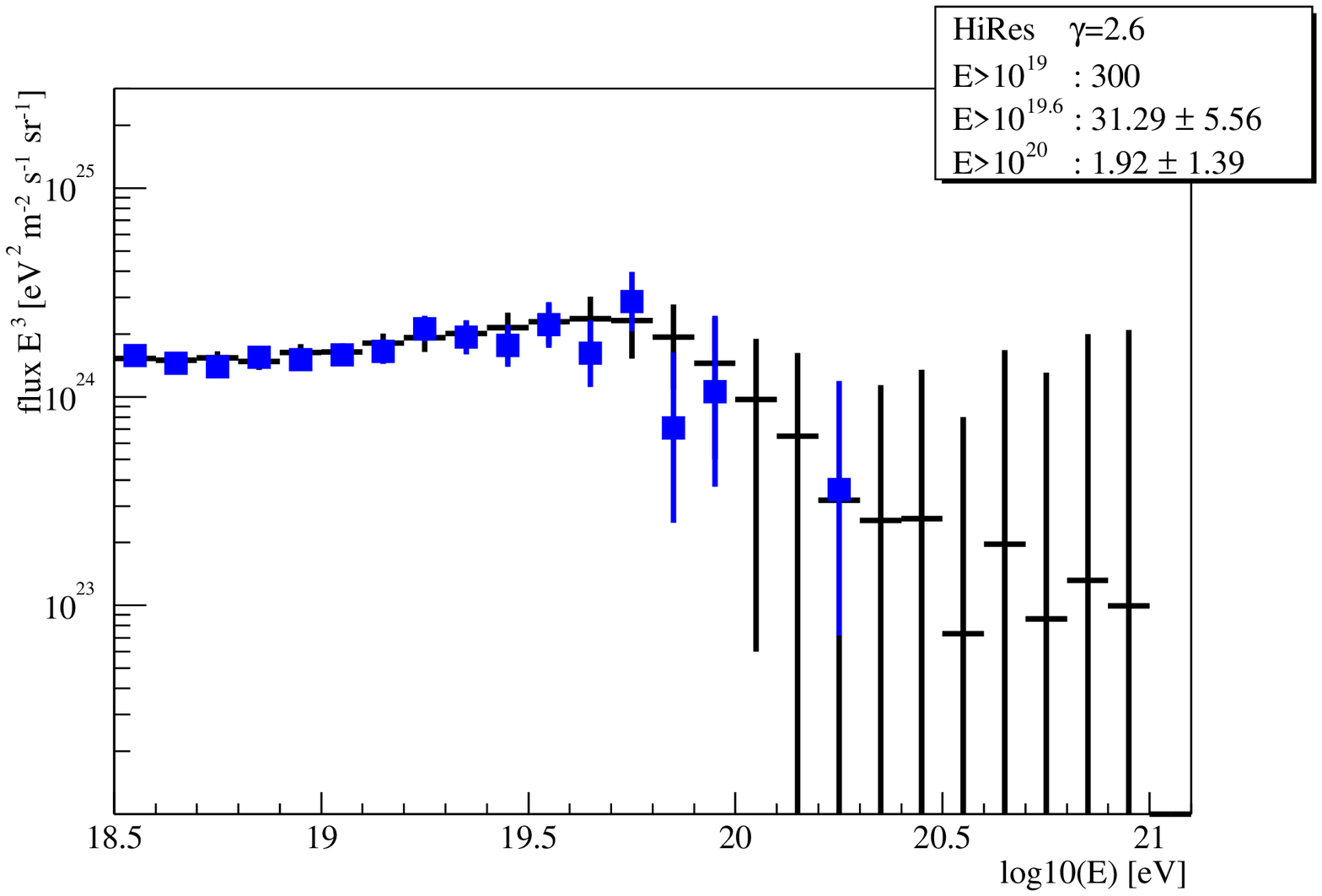}
\includegraphics[width=0.7\textwidth]{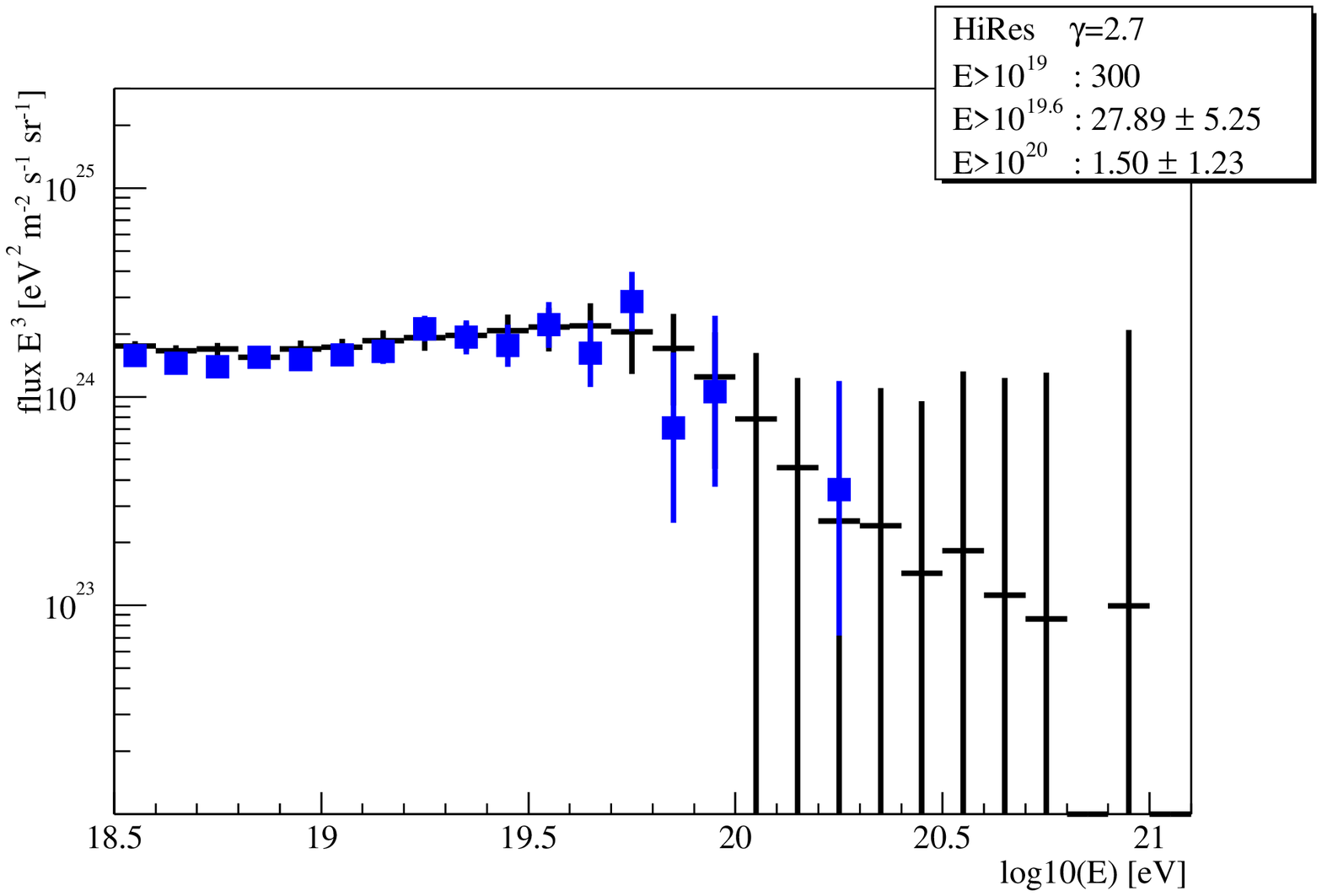}
\caption{Simulations of HiRes statistics with injection spectra
$E^{-2.6}$ (upper plot) and $E^{-2.7}$ (lower plot). The crosses with
error bars are the results of simulations, while the squares
are the HiRes data.}\label{fig:hires_2528}
\end{center}
\end{figure}

A more graphical representation of the uncertainties involved are
displayed in Figs. \ref{fig:agasa_2627} for AGASA and
\ref{fig:hires_2528} for HiRes, for two choices of the injection spectrum.
These plots show clearly the low level of significance that the detections
above $E_{GZK}$ have with low statistics.  The
large error bars that are generated by our simulations at the high
energy end of the spectrum are mainly due to
the stochastic nature of the process of photo-pion production: in some
realizations some energy bins are populated by a few events, while in
others the few particles in the same energy bin do not
produce a pion and get to the observer unaffected. The large
fluctuations are unavoidable with the extremely small statistics available
with present experiments. On the other hand, the error bars at lower energies
are minuscule, so that the two data sets (AGASA and HiResI) cannot be
considered to be two different realizations of the same phenomenon. Instead,
systematic errors in at least one if not both experiments are needed to
explain the discrepancies at lower energies.

Taking into account the (theoretical) error bars in the analysis makes
the significance of the presence or absence of the GZK feature much weaker
than the consistency checks shown in Tables  \ref{tab:discrepanza_agasa} and
\ref{tab:discrepanza_hires}. In order to estimate this effect, we
proceed in the following
way: we calculate the expected number of events above some threshold with
its corresponding standard deviation ($\sigma_{sim}$), as determined by the
fluctuations in the flux simulation. The observed number of events above the
same threshold is also known with the error bar $\sigma_{obs}$. The discrepancy
between the two is now calculated using the error
$\sigma_{tot}=(\sigma_{sim}^2+
\sigma_{obs}^2)^{1/2}$. Our results are summarized in Tables
\ref {tab:discrepanza_agasa} for AGASA and
\ref {tab:discrepanza_hires} for HiRes. The numbers with
error bars are the simulated expectations, while the discrepancy between
simulations and observations, calculated as described above is
in square brackets, in units of $\sigma_{tot}$.
It becomes clear that the effective discrepancy between predictions
and the AGASA data is at the level of $2.1-2.5\sigma$.
Therefore a definite answer to the question of whether the GZK feature
is there or not awaits a significant improvement in statistics at high
energies.

As seen in Fig. \ref{fig:agasahires}, the difference between the AGASA
and HiResI spectra is not only in the presence or absence
of the GZK feature: the two spectra, when multiplied by $E^3$, are
systematically shifted by about a factor of two. This shift suggests that
there may be a systematic error either in the energy or the flux determination
of at least one of the two experiments. Possible systematic effects have
been discussed in \cite{astro0209422} for the AGASA collaboration and in
\cite{HIRES2} for HiResI. At the lower end of the energy range the offset may
be due to the notoriously difficult determination of the AGASA aperture at
threshold while the discrepancies at the highest energies is unclear at the
moment. In any case, a systematic error of $\sim 15\%$ in the energy
determination is well within the limits that are allowed by the analysis of
systematic errors carried out by both collaborations.

In order to illustrate the difficulty in determining the
existence of the GZK feature in the observed data in the
presence of systematic errors, we split the energy gap by assuming that
the energies as determined by the AGASA collaboration
are overestimated by $15\%$, while the HiRes energies
are underestimated by the same factor. The number of events
above an energy threshold is again reported in Table
\ref{tab:data_shifted}, in the column labelled $15\%$.
In this case the total number of events above $10^{19}$ eV
is reduced for AGASA from 866 to 651, while for HiResI
it is enhanced from 300 to 367. We ran our simulations
with these new numbers of events and repeat the statistical analysis
described above.  The values of $\chi^2$ obtained in this case are reported in
Table  \ref{tab:chiq_shift}.
\begin{table}
\begin{center}
\caption{$\chi^2$ for AGASA and HiResI in which a correction for a systematic
$15\%$ overestimate of the energies has been assumed for AGASA and a
$15\%$ underestimate of the energies has been assumed for HiResI.}
\begin{tabular}{l|ccc|ccc}
&\multicolumn{3}{c|}{AGASA}&\multicolumn{3}{c}{HiResI}\\
$\gamma$    &   $\chi^2_{18.6}(15)$ &   $\chi^2_{19}(11)$   &
$\chi^2_{19.6}(5)$ &   $\chi^2_{18.6}(14)$ &   $\chi^2_{19}(10)$   &
$\chi^2_{19.6}(4)$\\
\hline
2.3 & 505     & 18       & 12       & 79       & 13    & 7\\
2.4 & 351     & 13       & 8.5      & 40       & 7     & 4\\
2.5 & 188     & {\bf 9}  & {\bf 5.6}& 13       & 3.7     & 2.0\\
2.6 & 54      & {\bf 9}  & {\bf 5.6}& {\bf 6}& {\bf 2.0} & {\bf 1.1}\\
2.7 & {\bf 20}& 11       & 6.4      & 23       & 3.1     & 1.4\\
2.8 & 54      & 15       & 7.2      & 94       & 6     & 2.4\\
2.9 & 145     & 20       & 9.1      & 176      & 10    & 4\\
\hline
\end{tabular}
\label{tab:chiq_shift}
\end{center}
\end{table}

For AGASA, the best fit injection spectrum is now between $E^{-2.5}$
and $E^{-2.6}$ above $10^{19}$ eV and above $10^{19.6}$ eV (the $\chi^2$
values are identical). For the HiRes data,
the best fit injection spectrum is $E^{-2.6}$ for the whole set
of data, independent of the threshold. It is interesting to note that the
best fit injection spectrum appears much more stable after the
correction of the $15\%$ systematics has been carried out. Moreover,
the best fit injection spectra as derived for each experiment
independently coincides for the corrected data unlike the uncorrected case.
This suggests that combined systematic errors in the energy determination
at the $\sim$ 30\% level may in fact be present.

The expected numbers of events with energy above $10^{19.6}$ eV
and above $10^{20}$ eV with the deviation from the data are
reported in Tables \ref{tab:discrepanza_agasa085} and
\ref{tab:discrepanza_hires115}: while HiResI data remain in
agreement with the prediction of a GZK feature, the AGASA
data seem to depart from such prediction but only at the
level of $\sim 1.8\sigma$. The systematics in the energy
determination further decreased  the significance
of the GZK feature, such that the AGASA and
HiResI data are in fact only less than $2\sigma$ away from each other.
\begin{table}
\begin{center}
\caption{Number of events expected above $E_{th}$ (eV) for AGASA energies
decreased systematically by $15\%$. In parenthesis is the number of
standard deviations between observations and simulations, $\sigma$. In
square brackets are the discrepancies calculated in units of
$\sigma_{tot}$.}
\begin{tabular}{c|c|c|c}
${E_{th}}$ & $\gamma=2.5$ & $\gamma=2.6$ & $\gamma=2.7$ \\
\hline
$10^{19.6}$ & \dsc{49}{6.9}{+0.2}{+0.1} & \dsc{43}{6.5}{+0.8}{+0.5} &
\dsc{39}{6.1}{+1.3}{+1.0} \\ $10^{20}$   & \dsc{2.6}{1.6}{+1.7}{+1.4} &
\dsc{2.3}{1.5}{+1.8}{+1.5} & \dsc{1.8}{1.4}{+2.0}{+1.7} \\
\hline
\end{tabular}
\label{tab:discrepanza_agasa085}
\end{center}
\end{table}

\begin{table}
\begin{center}
\caption{Number of events expected above $E_{th}$ (eV) for HiResI energies
increased systematically by $15\%$. In parenthesis is the number of
standard deviations between  observations and simulations, $\sigma$. In
square brackets are the discrepancies calculated in units of
$\sigma_{tot}$.}
\begin{tabular}{c|c|c}
${E_{th}}$&$\gamma=2.5$ & $\gamma=2.6$ \\
\hline
$10^{19.6}$ & \dsc{43}{6.3}{-0.6}{-0.4} & \dsc{38}{6.0}{+0.1}{+0.1} \\
$10^{20}$   & \dsc{2.8}{1.7}{-0.4}{-0.3} & \dsc{2.3}{1.5}{-0.1}{-0.1} \\
\hline
\end{tabular}
\label{tab:discrepanza_hires115}
\end{center}
\end{table}

\begin{figure}
\begin{center}
\includegraphics[width=0.7\textwidth]{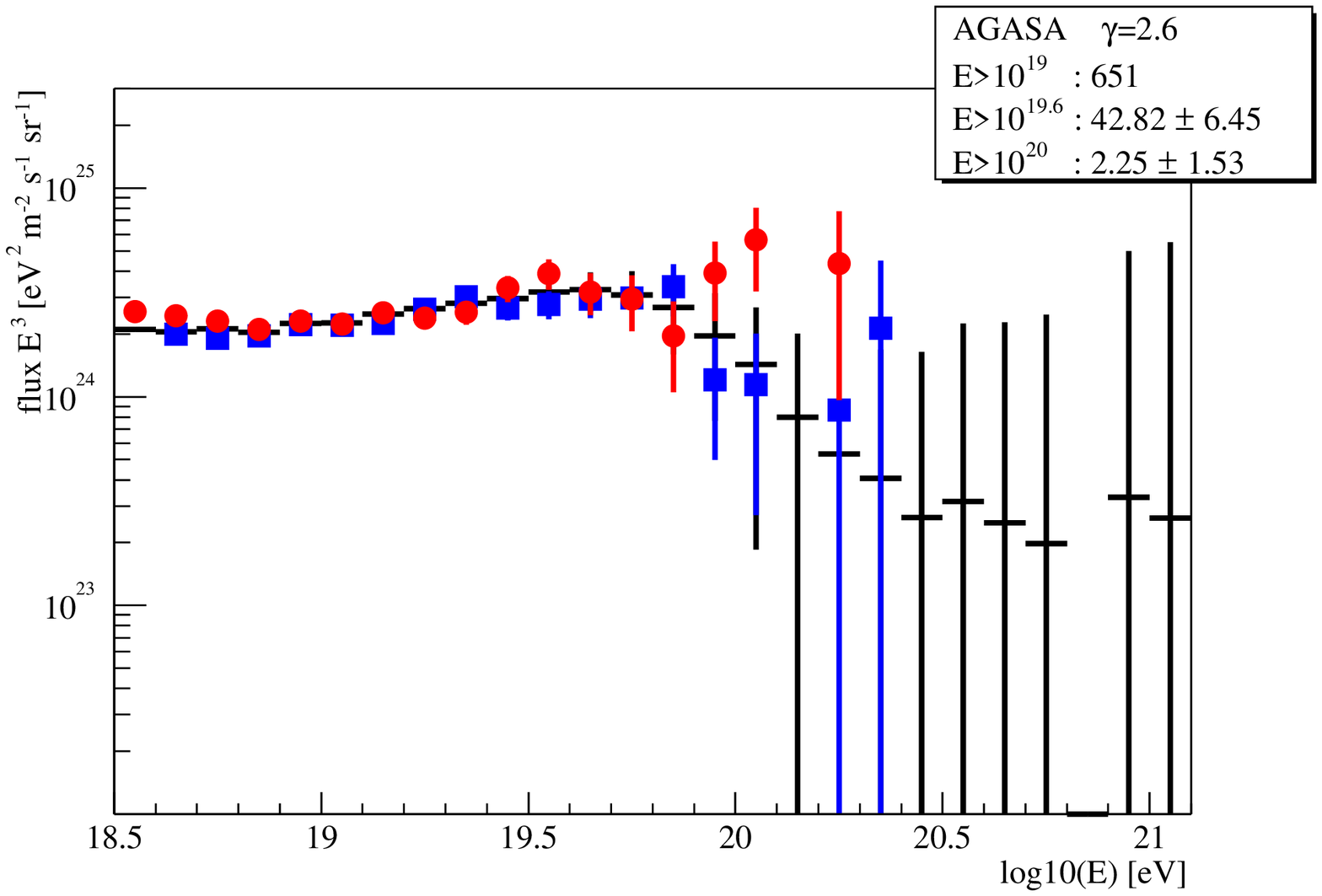}
\includegraphics[width=0.7\textwidth]{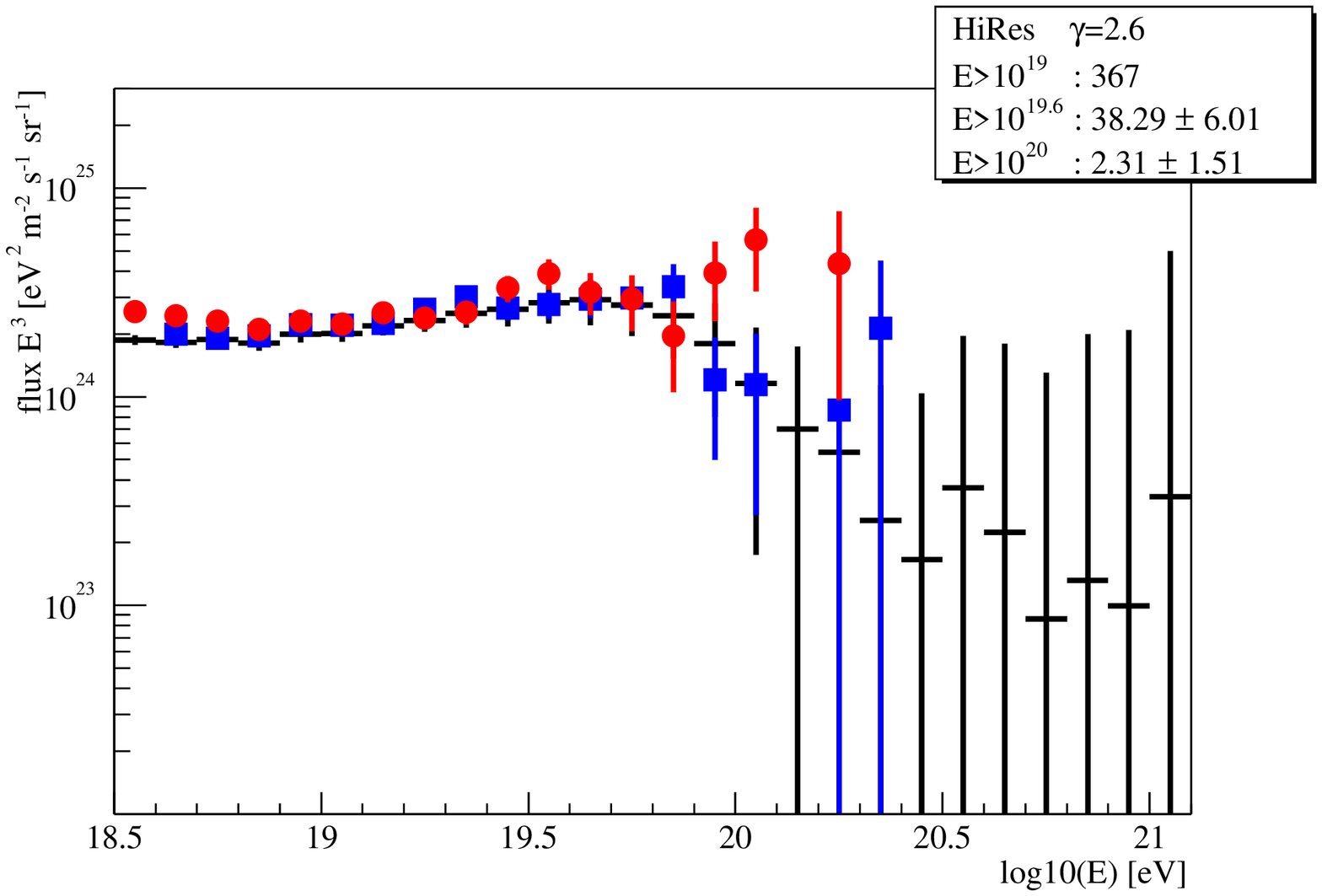}
\caption{Simulated spectra for the best fit injection spectrum with
$\gamma=2.6$. Upper panel shows simulations for the AGASA event statistics
after correcting for energy by an overall shift of -15\%. The lower panel
shows  the fluctuations expected for the event statistics of HiResI after
shifting HiResI energies by +15\%.
The shifted data for AGASA (grey circles) and  HiResI (dark squares) are shown
in both panels.}
\label{fig:26big}
\end{center}
\end{figure}

We can use the same procedure illustrated above to
evaluate the effect of the error bars in the simulated data compared to the
data corrected by 15\%. The results are reported in square brackets in Tables
\ref {tab:discrepanza_agasa085} (for AGASA) and \ref {tab:discrepanza_hires115}
(for HiRes), showing that the effective discrepancy between expectations (with
uncertainties due to discrete energy losses and cosmic variance) and AGASA
data is even smaller, only at the $1.5\sigma$ level.  In Fig.
\ref{fig:26big}, we plot the simulated spectra for injection spectrum
$E^{-2.6}$ and compare them to observations of AGASA (upper plot) and HiRes
(lower plot).

\section{Conclusions}

We considered the statistical significance of the UHECR spectra
measured by the two largest experiments in operation, AGASA and HiRes. The
spectrum released by the HiResI collaboration  seems to suggest the presence
of a GZK feature. This has generated claims that the GZK
cutoff has been detected, reinforced by data from older experiments
\cite{waxmanb}. However, no evidence for
such a feature has been  found in the AGASA experiment.  We compared the
data with theoretical predictions for the propagation of UHECRs on
cosmological distances with the  help of numerical simulations.
We find that the very low statistics of the presently available data
hinders any statistically significant claim for either detection or the
lack of the GZK feature.

A comparison of the spectra obtained from AGASA and
HiResI shows a systematic shift of the two data sets, which may be
interpreted as a systematic error in the relative energy determination
of about $30\%$. If no correction for this systematic shift is carried
out, the AGASA data are best fit by an injection spectrum $E^{-\gamma}$
with $\gamma=2.6$ for energies above $10^{19}$ eV. The fit steepens to
$\gamma=2.8$ when considering events down to $10^{18.5}$ eV. For HiResI,
the best fit is between $\gamma=2.7$ and $\gamma=2.8$ for events with
energy above $10^{19}$
eV and $\gamma=2.6$ for  events above
$10^{18.5}$ eV. With these best fits to the injection spectrum the
AGASA data depart from the prediction of a GZK feature by $2.6\sigma$
for $\gamma=2.6$. The HiRes data
are fully compatible with the prediction of a GZK feature in the
cosmic ray spectrum. The fit to the data with energy above $10^{19}$ eV
is probably less susceptible to contamination by a possible galactic flux. In
this case the AGASA departure from the expected GZK feature is, as stressed
above, at the level of about $2.6\sigma$.
Taking into account the uncertainties derived from the simulations,
attributed to the discreteness of the photo-pion production and to
cosmic variance, this discrepancy becomes even less significant ($\sim
2.3\sigma$). It is clear that, if confirmed by future experiments with
much larger statistics, the increase in flux  relative to the GZK
prediction hinted by AGASA would be of great interest. This may signal the
presence of a new component at the highest energies that compensates for the
expected suppression due to photo-pion production, or the effect of new
physics in particle interactions (for instance the violation of Lorentz
invariance or new neutrino interactions).

Identifying the cause of the systematic energy and/or flux shift between
the AGASA and the HiRes spectra is crucial for understanding the nature
of UHECRs. This discrepancy has stimulated a number of efforts  to search
for the source of these systematic errors including the construction of
hybrid detectors, such as Auger, that utilize both ground arrays and
fluorescence detectors. A possible overestimate of the AGASA energies by
$15\%$ and a corresponding underestimate of the HiRes energies by the same
amount would in fact bring the two data sets in agreement in the
region of energies below $10^{20}$ eV. In this case both experiments are
consistent with a GZK feature with large error bars. The AGASA excess
is at the level of  $1.7\sigma$ ($1.4\sigma$ if the observational
uncertainties are also taken into account). Interestingly
enough, the correction by $15\%$ in the error determination implies
that the best fit injection spectrum becomes basically the same
for both experiments ($E^{-2.6}$).

\begin{figure}
\begin{center}
\includegraphics[width=0.7\textwidth]{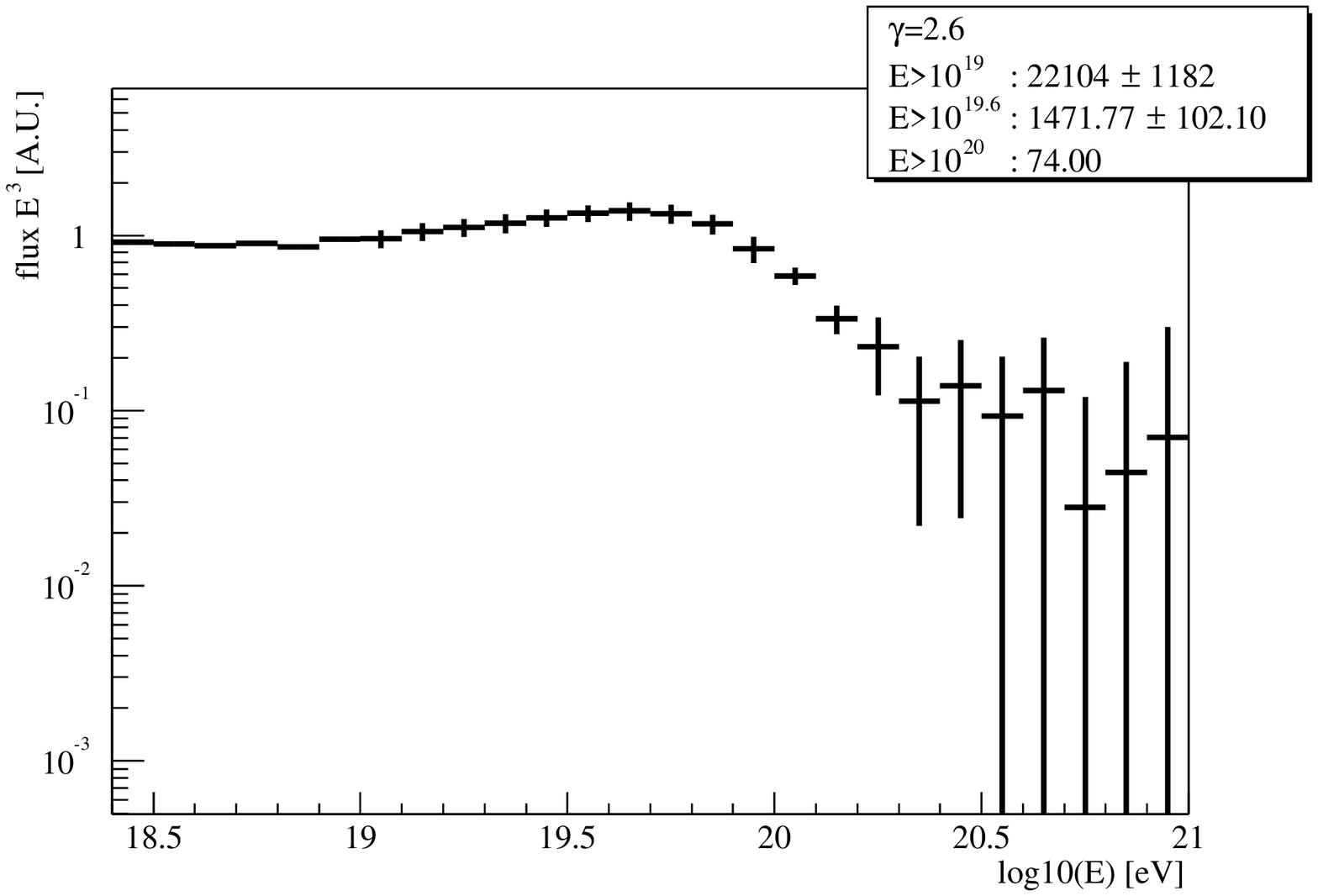}
\includegraphics[width=0.7\textwidth]{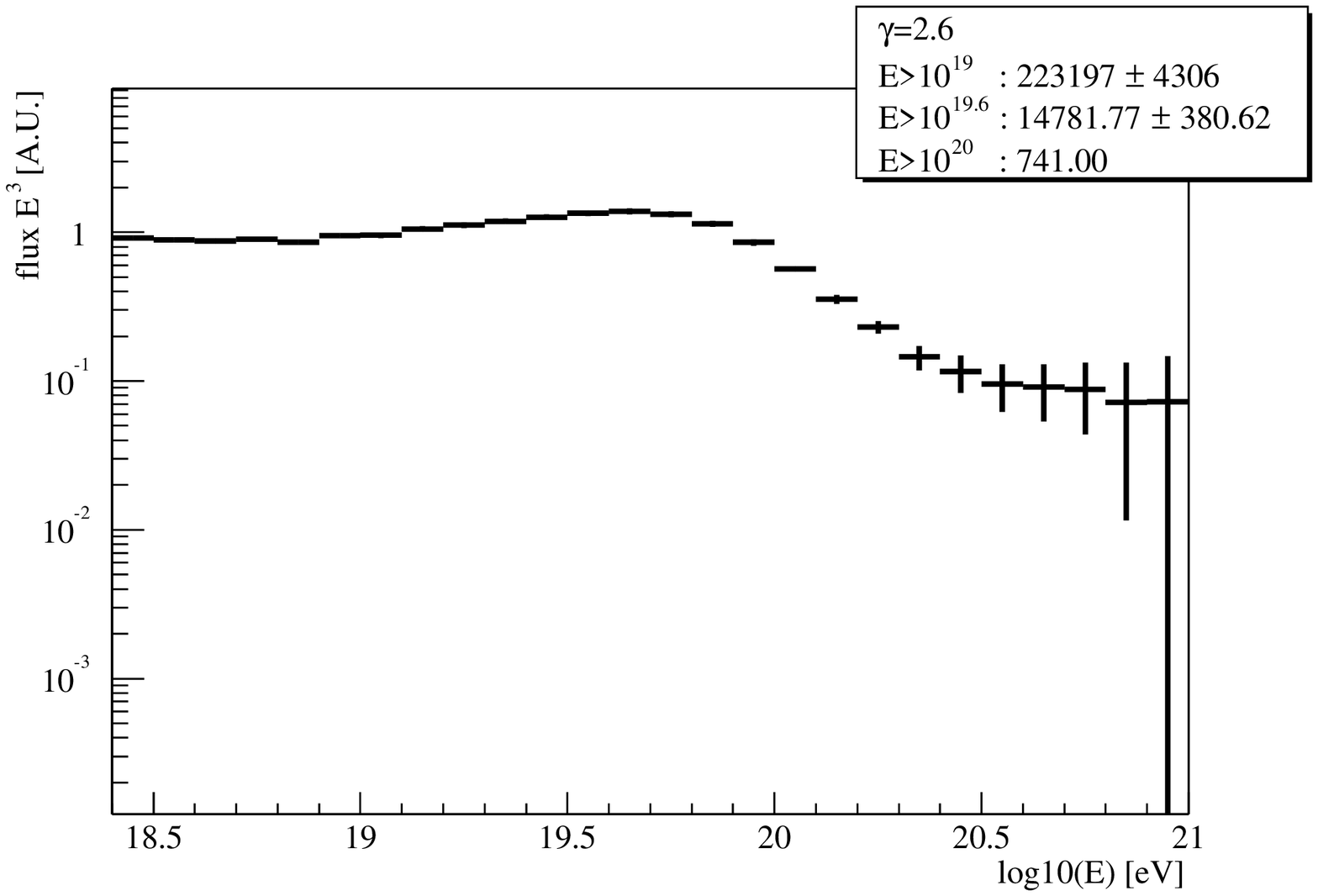}
\caption{Predicted spectra and error bars for 3 years of operation of
Auger (upper plot) and EUSO (lower plot).}
\label{fig:newgen}
\end{center}
\end{figure}

With the low statistical significance of either the excess flux seen by
AGASA or the discrepancies between AGASA and HiResI, it is inaccurate to
claim either the detection of the GZK feature or the extension of the
UHECR spectrum beyond $E_{GZK}$ at this point in time. A new generation
of experiments is needed to finally give a clear answer to
this question. In Fig. \ref{fig:newgen} we report the simulated spectra
that should be achieved in 3 years of operation of Auger (upper panel)
and EUSO (lower panel). The error bars reflect the fluctuations expected
in these high statistics experiments for the case of injection spectrum
$E^{-2.6}$. (Note that the energy threshold for detection by EUSO is not
yet clear.) It is clear that the energy region where statistical
fluctuations dominate the spectrum is moved to $\sim 10^{20.6}$ eV for
Auger, allowing a clear identification of the GZK feature. The {\it
fluctuations} dominated region stands beyond $10^{21}$ eV for EUSO.

\section*{Acknowledgments} We thank Douglas Bergman for providing the number of
events in different bins for the HiRes results. We also thank Venya Berezinsky,
James Cronin, Masahiro Teshima, Mario Vietri and Alan Watson for a number of
helpful comments. This work was supported in part by the NSF through grant
AST-0071235 and DOE grant DE-FG0291-ER40606 at the University of Chicago.


\end{document}